# Development of Design for the STS Extraction Magnet System

V. Chernenok*, S. Cheban, D. J Harding, V. Kashikhin, T. Strauss, B. Szabo, Fermi National Accelerator Laboratory
M. Allitt, L. Boyd, Oak Ridge National Laboratory

*Abstract*—The Oak Ridge National Laboratory STS Project will enhance the Spallation Neutron Source by adding a new neutron source. The upgrade includes a 30% increase in beam energy and a 50% boost in beam current, doubling the accelerator's power capability to 2.8 MW. The Ring-to-Second-Target Beam Transport (RTST) system is vital in directing high-energy proton beams to the new second target station. New magnets have to fit into a restricted space within the existing beamline to add the new extraction line. New, fast kicker magnets (Pulsed Dipole), a focusing quad with an aperture for both beamlines (Large Quad) and the replacement of existing 21Q40 Quads with 'Narrow Quads' having identical fields as the existing design are Fermilab's responsibility. The design of these magnets poses unique challenges, as in addition to the high requirements on the quality of magnetic fields, they are subject to major restrictions related to their dimensions.

## I. INTRODUCTION

THE Oak Ridge National Laboratory (ORNL) is enhancing the Spallation Neutron Source (SNS) [1-6] through the development of a Second Target Station (STS) [6]. This initiative aims to address critical challenges in materials research by providing intense, cold neutron beams for exploring complex materials. The STS will complement the existing First Target Station (FTS) [7] and the High Flux Isotope Reactor (HFIR), significantly expanding the experimental capabilities of the SNS.

Central to this upgrade is the Ring-to-Second-Target Beam Transport (RTST) system, designed to reliably and precisely direct high-energy proton beams to the new target. The RTST system relies on advanced magnetic components to ensure proper beam extraction, steering, and focusing while accommodating stringent spatial and operational constraints, shown in Fig. 1 is the extraction section.

Within this work, Fermilab is responsible for developing three types of magnets for the RTST system: Pulsed Dipole, Large Aperture Quadrupole, and Narrow Quadrupole. The RTST system will incorporate four Pulsed Dipole magnets for beam deflection, along with one Large Aperture Quadrupole and one Narrow Quadrupole magnet to provide the necessary beam focusing and alignment. Each magnet type has been engineered to meet strict performance and dimensional requirements, ensuring compatibility with the spatial and operational constraints of the upgraded SNS accelerator system.

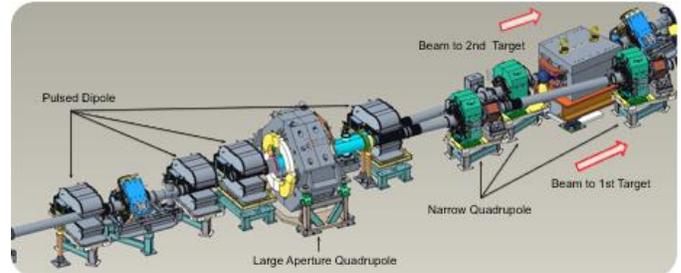

**Fig. 1.** Extraction beamline section of the new RTST system

## II. MAGNETIC PERFORMANCE REQUIREMENTS

Base on Second target Station Beam Line Optics documents developed by specialists from the Oak Ridge National Laboratory, the main parameters of the magnets are presented in Table 1. The Narrow Quad and Large Quad operate in DC mode, however for the Pulsed Dipole the peak magnetic field shall exhibit pulse-to-pulse repeatability within ± 0.1 % and achieve field stabilization within a few milliseconds after the end of the ramp.

Achieving the required magnetic performance, thermal stability, and mechanical integrity involved extensive use of 3D finite-element modeling tools, particularly the OPERA3D simulation software [8].

For all magnets in the extraction region, the field gradient uniformity and integrated field accuracy are required to be within ±0.1 % throughout the specified good field region. This uniformity criterion ensures consistent beam focusing and deflection performance across the entire aperture.

TABLE 1
MAGNETS PARAMETERS

|  |  | Pulsed Dipole | Large Aperture Quad. | Narrow Quad. |
|---|---|---|---|---|
| **Parameters** | **Units** | Value | | |
| Gap, Aperture | mm | 250 | 400 | 209 |
| Total Current/per coil | A | 38,543 | 50,706 | 27,400 |

*V. Chernenok (corr. Author: vchernen@fnal.gov), S. Cheban scheban@fnal.gov, David J Harding harding@fnal.gov, V. Kashikhin kash@fnal.gov, T. Strauss strauss@fnal.gov, B. Szabo bszabo@fnal.gov, are with the Fermi National Accelerator Laboratory, P.O. 500, Batavia, IL 60510 USA

Allitt, Michael allittml@ornl.gov, Boyd Larry boydlm@ornl.gov, are with the Oak Ridge National Laboratory, P.O. Box 2008, 1 Bethel Valley Road Oak Ridge, TN 37831-6324

TABLE 2 (continued)

| Peak current | A | 1,606 | 1,538 | 978 |
|---|---|---|---|---|
| Gap center field | T | 0.188 | N/A | N/A |
| Center field gradient | T/m | N/A | 3.185 | 2.955 |
| Integrated field gradient | T | N/A | 2.806 | 2.956 |
| Integrated field | T·m | 0.166 | N/A | N/A |
| Core length | mm | 500 | 700 | 400 |
| Effective length | mm | 883 | 881 | 487 |
| Good field Region | mm | ± 45 (H) ± 57 (V) | Radius ≥ 176 | Radius ≥ 70 |

III. LARGE APERTURE QUADRUPOLE MAGNET DESIGN

The large aperture quadrupole operated in DC mode. The magnet core is made from low-carbon solid steel pieces. Main parts of magnet core have flux density below 1 T besides pole edges (Fig. 2)

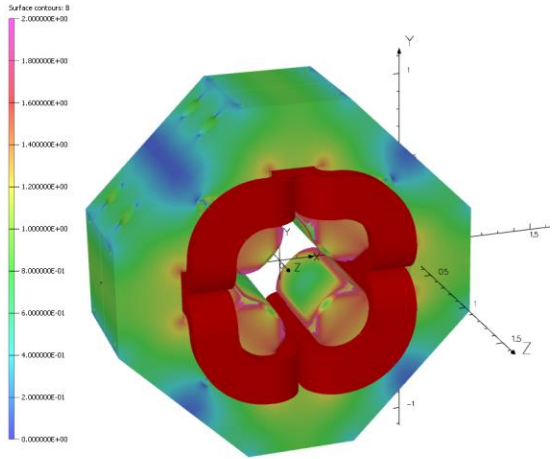

**Fig. 2.** Large aperture quadrupole flux density.

The large quad design needed to fulfill three main criteria, fit the existing tunnel location, be splitable in operation to allow coil replacement and allow clearance and good field region for the separated beam pipes of the primary and secondary beamline. Fig. 3 illustrates the final assembly model of a large aperture quadrupole magnet with a magnet stand. The magnet bore diameter is 400 mm.

The end-to-end dimensions of the finished assembly are 1090 mm (length) × 2230 mm (width) × 1980 mm (height). The magnet weight 17 metric tons. The model includes electrical and hydraulic connections for the coil packs. Each core quarter of the magnet is constructed from AISI 1006 steel, weighing approximately 4 metric tons.

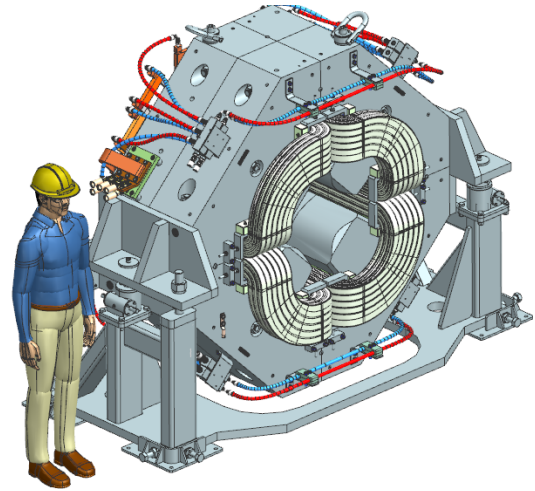

**Fig. 3.** Large aperture quadrupole magnet assembly and stand.

The coils of all three magnet types will be impregnated using vacuum pressure impregnation using radiation-resistant epoxy resin reinforced with fiberglass insulation. These resin systems are widely used in accelerator magnets exposed to significant doses of ionizing radiation and retain their dielectric and mechanical properties.

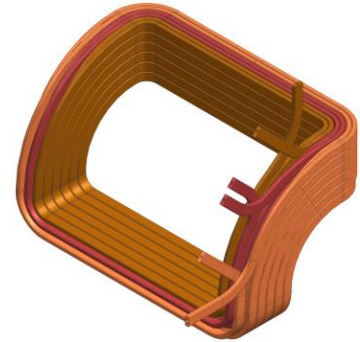

**Fig. 4.** Large aperture quadrupole coil assembly.

The magnet features four identical coil assemblies, each composed of three coils (Fig. 4). The coils are wound from a 15.9 mm × 38.1 mm Luvata copper conductor, incorporating a 3/8-inch cooling channel. The total copper mass of each coil assembly is approximately 550 kilograms, excluding insulation.

Hydraulic cooling is facilitated by a set of Parker Series 56 stainless steel crimp-style fittings and Parker Parflex 518D hoses. The water-cooling system includes three separate flow paths per coil assembly, with water flow directed to the inner conductor and returned from the outer conductor to optimize thermal management.

To prevent thermal runaway during operation, Sensata 4344 thermal sensors are affixed to the insulated conductor near the water outlets. These sensors monitor the coil's thermal state and provide an interlock to the power supply. The system is designed to maintain a temperature gradient of less than 5°C across the water flow, with a Reynolds number of approximately 25,000, ensuring turbulent flow.

To facilitate precise alignment during installation and operation, the magnet is equipped with 64 milling surfaces for fiducial. These monuments enable alignment accuracy within 100 microns in the x and y axes and 1 milliradian in yaw, pitch, and roll.

## IV. PULSED DIPOLE

The dipole magnet operates in a pulsed current mode. The pulses, 30 ms long, have a repetition rate of 15 Hz. This drives the design to the laminated low-carbon steel core.

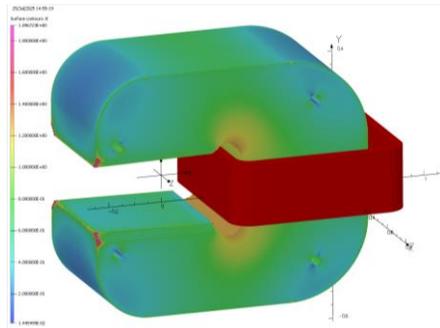

**Fig. 5.** Magnet flux density distribution.

The pulsed dipoles act as fast kicker magnets for the secondary beamline extraction. The dynamic performance requirements for the pulsed dipole specify a maximum field lag < 0.1 %, field stabilization within < 2 ms after the current ramp, and pulse-to-pulse repeatability better than ±0.1 %. AC transient simulations show that eddy-current–induced field distortion in the magnet aperture remains below 0.02 %. The laminated yoke effectively minimizes magnetic diffusion time, and the Inconel beam pipe introduces negligible additional delay or field perturbation at the 15 Hz operating frequency.

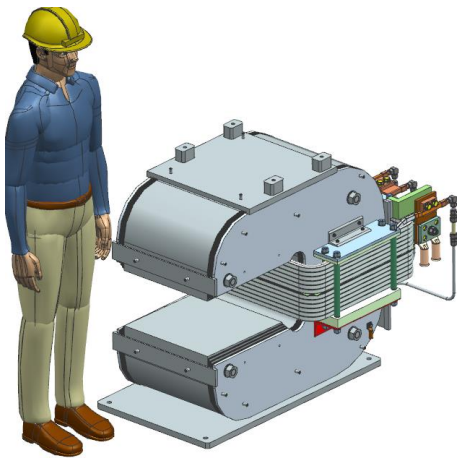

**Fig. 6.** Pulsed dipole assembly.

Fig. 6 illustrates the final assembly model of the pulsed dipole magnet. The magnet's overall dimensions are 780 mm (length) × 1550 mm (width) × 975 mm (height), with a total weight of approximately 3.5 metric tons. The magnet core comprises of 1000 precisely cut laminations to ensure optimal magnetic field uniformity and efficiency. The current pulses have the 15 Hz equivalent frequency, and it was chosen the 0.5 mm lamination thickness which is well below the ~1 mm skin depth. Structural integrity is further enhanced by stainless steel reinforcement plates welded around the assembly. The magnet includes two identical coils (Fig. 7), each weighing 45 kg (copper only). Each coil is wound from a single continuous length of copper conductor with cross-sectional dimensions of 12 mm × 18 mm and features an integrated 8 mm cooling channel.

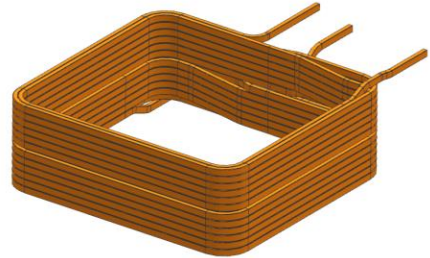

**Fig. 7.** Pulsed dipole coils

Thermal stability is maintained with a water circuit temperature rise limited to 6°C. Each coil is equipped with two thermal switches, directly connected to the power supply, providing protection against overheating.

To facilitate precise alignment during installation and operation, the magnet is equipped with 24 welded fiducial monuments. These monuments enable alignment accuracy within 100 microns in the X and Y-axis and 1 milliradian in yaw, pitch, and roll. A wood prototype was built to test assembly techniques, material handling and fabrication procedures, the lessons learned helped improve the design.

## V. NARROW QUADRUPOLE

The special quadrupole magnet is designed to fit in the narrow space between two beam lines. The old quadrupole tests showed the large iron core saturation. To reduce saturation, the top and bottom core thicknesses were increased. In this case, the magnetic flux mostly goes through thick parts forming so named Figure-8 quadrupole. The resulting flux density is shown in Figure 8.

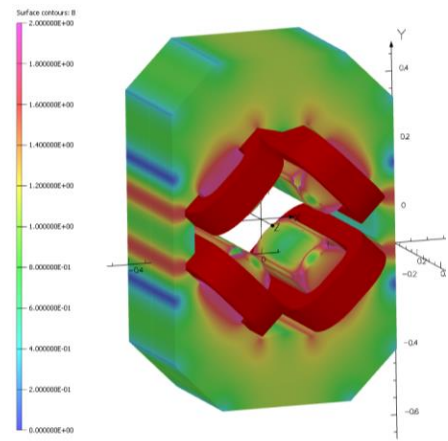

**Fig. 8.** Narrow quadrupole flux density

Fig. 9 illustrates the final assembly model of the Narrow quadrupole. The magnet's overall dimensions are 776 mm (length) × 1023 mm (width) × 1040 mm (height), with a total weight of approximately 1.9 metric tons.

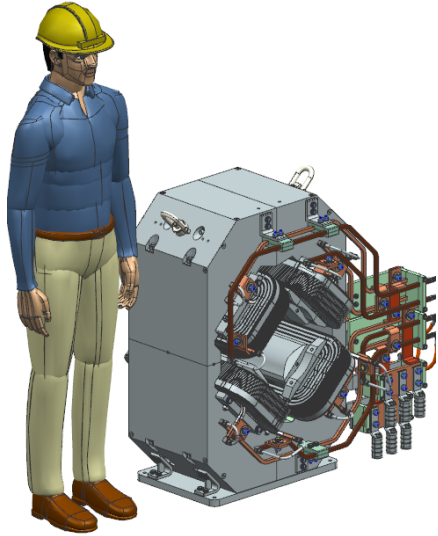

**Fig. 9.** Narrow quadrupole assembly

The magnetic core consists of four identical segments made of AISI 1006 low carbon steel, each weighing approximately 376 kg. The assembly is secured using M16 tie rods, highlighted in Fig. 10, that hold the quarter-cores segments together. This type of connection design is used for the first time and allows for a radical reduction in the width of the magnet.

A through hole passes through the thin side of the quarter core, through which the quarters will be connected to each other. This is one of the most difficult challenges on this project, however mechanical analysis ANSYS [9] model revealed no significant load in this area, with stress measured at 32 MPa.

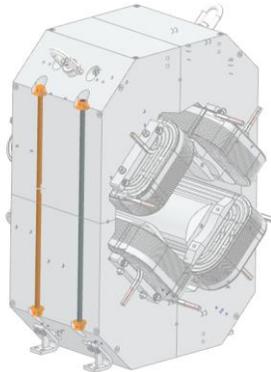

**Fig. 10.** Rods connection

The optimized pole profile for the narrow quadrupole, as shown in Fig. 11, was based on the previously magnet design, which demonstrated excellent field quality during years of reliable operation. The final shape retained the same basic contour, with minor flat regions added to improve linearity. Magnetic simulations confirmed that all higher-order field harmonics remain below 0.02 % at the reference radius of 75 mm, well within the < 0.1 % field quality requirement.

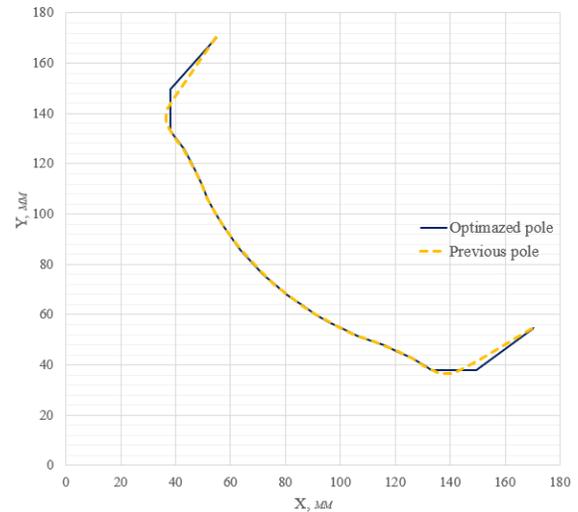

**Fig. 11.** Optimized pole profiles

The quadrupole consists of four coils, one per pole, each forming a parallel cooling branch. Each coil has 28 turns wound from a 12.7 mm square conductor with a 6.4 mm central cooling hole, using fiberglass sleeve insulation applied during winding. Two trim circuits with 2 and 14 turns, respectively, are made from AWG 6 solid conductor. The coil assembly weighs approximately 63 kg, including insulation. Operating at 978 A, the system dissipates 5.6 kW, with a current density of 7.65 A/mm² and a temperature rise limit of 24.5 °C (77.9°F). This coil is a replica design of the previously fabricated 21Q40 to allow compatibility and new spare production over the lifetime of the new STS beamline.

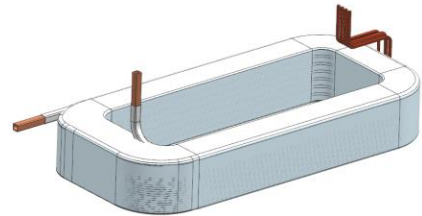

**Fig. 12.** Narrow quadrupole coils

## VI. CONCLUSION

The RTST magnet designs meet the specified extraction region requirements. They have been fully reviewed and are ready for fabrication, including the necessary tooling design. Each magnet design fits its specific lattice requirements for the future beam line, as well as the serviceability over the life of the facility. The next step is to fabricate prototypes and perform magnetic measurements to verify compliance before the full production.

## ACKNOWLEDGMENT

The authors would like to express their sincere gratitude to the U.S. Department of Energy Office of Science, as well as the dedicated teams and management at Oak Ridge National Laboratory and Fermilab, for supporting this collaborative effort between two U.S. National Laboratories.

MT29-Thu-Af-Po.01-03                                                                 5## REFERENCES

[1] Spallation Neutron Source (SNS) – Oak Ridge National Laboratory, 2025, [Online]. Available: https://neutrons.ornl.gov/sns

[2] S. Henderson, "Status of the SNS beam power upgrade project," in Proc. EPAC, Edingburgh, Scotland, 2006, pp. 345–347.

[3] J. G. Wang, "Magnetic field distribution of injection chicane dipoles in spallation neutron source accumulator ring," Phys. Rev. Special Top.-Accel. Beams, vol. 9, 2006, Art. no. 012401.

[4] J. G. Wang, "Performance improvement of an extraction lambertson septum magnet in the spallation neutron source accumulator ring," Phys. Rev. Special Top.-Accel. Beams, vol. 12, 2009, Art. no. 042402.

[5] P. Wanderer, "The SNS ring dipole magnetic field quality," in Proc. EPAC, Paris, France, 2002, pp. 1317–1319.

[6] Second Target Station Project – ORNL, 2025, [Online]. Available: https://neutrons.ornl.gov/sts

[7] V. Kashikhin, J. Amann, N. Evans, D.J. Harding, J. Holmes, M. Plum, D. Pomella, "Magnetic Designs of New First Target Beamline Magnets for the ORNL SNS Upgrade," in IEEE Trans. Appl. Supercond., vol. 32, no. 6, Sep. 2022, Art. no. 4001604, pp. 1–4.

[8] "Opera Simulation Software," Dassault Systems UK Ltd 1984-2021, [Online]. Available: www.3ds.com/products-services/simulia/products/opera.

[9] "ANSYS Simulation Software," 2025 [Online]. Available: https://www.ansys.com/contact-us.